\documentstyle{aipproc}
\begin{document}
\begin{flushright}
KUNS-1694\\
astro-ph/0010559\\
\today
\end{flushright}
\title{Vacuum Selection by Collapsing}
\author{Nobuhiro Maekawa}
\address{Department of Physics, Kyoto University,\\ 
Kyoto 606-8502, Japan}
\maketitle
\begin{abstract}
In this talk, we discuss the possibility that the vacuum is dynamically 
determined in the history of the universe. The point is that some of 
the bubbles with a certain vacuum shrink by the evolution of the 
universe via gravity and may become black holes. 
It is interesting that in many cases false vacua are favored in the 
context of cosmology. By using this argument, supersymmetry(SUSY), if 
it exists, can be broken cosmologically. 
This talk is based on Ref.\cite{maekawa}.
\end{abstract}

\section*{Introduction}
Our motivation of this work is to understand the elementary particle
physics, especially beyond the standard model. 
Usually elementary particle physicists expect to know the physics
beyond the standard model by the experiments on particle physics
(or for some people, by mathematical consistency). However, it may
happen that the past of our universe gives some clues to explain
some features of 
the rule of the nature(physics) as the past of human beings plays 
important roles in their behaviors. 
In this talk, we examine the
possibility that the cosmology selects theories of elementary particle
physics. Of course, it has been argued that the cosmology gives some
constraints to the theory of the elementary particle physics 
in order not to
break the success of the standard cosmology. For examples, the number of
massless neutrinos must be around three in order to realize the Big Bang
Nucleosynthesis. But such usual requirements are only constraints. Such
an argument never explain why the number of neutrinos is three. 
We discuss the possibility that the cosmology explains something on the 
elementary particle physics. The point is following.
The elementary particle theories determine the evolution of the universe. 
Some universes are expanding and other universes are shrinking. As the 
result, there are universes with various lifetimes.
It is natural to expect that our universe has a tendency to have the 
features which the universes with longer life have.

But as you know the universe is only one. What does many universes mean?
The answer in this talk is following.
In the phase transition, generally there are several local or global 
minima. At that time, the many bubbles with different vacua can exist. 
Different vacua mean different physics, and different physics leads to 
different evolution of the universe.
As the result, selection of the vacua occurs. 
Some bubbles may expand rapidly and dominate the universe, and other 
bubbles may shrink and disappear or become black holes.

There are lots of papers on the selection of the vacua
\cite{linde,izawa,banks}
 in the context 
of the inflation after the work on the chaotic inflation by Linde, but 
as long as I know, there are only a few papers on the 
selection of the vacua by recollapsing
\cite{banks}. Banks et al discussed the selection
only in the special case in the context of the string moduli problem.

Here we discuss the more general cases, in which
the vacuum selection by recollapsing happens,
because it is more important in the context of the selection of the 
theory. And we examine the features of the black holes which may appear 
as the Fossils.
In the last of this talk we try to apply this argument to the 
elementary particle physics, especially to SUSY model.

\section*{Vacuum selection}
In this section we argue on the possibility that collapsing bubbles and
expanding bubbles exist simultaneously, namely vacuum selection by 
recollapsing happens.

It is not easy to understand the evolution of the universe in phase 
transition, because the universe is essentially not homogeneous nor 
isotropic. Therefore we adopt the following simplifications.

The first assumption is that we use the Friedmann equation for the 
evolution of the bubble universes.
We expect that the size of the bubbles is so large, maybe the Hubble 
scale, that the space in the bubble is almost homogeneous and 
isotropic.

The second assumption is  that the phase transition does not change 
the energy density. This is only for the simplicity.

Under the first assumptions, we obtain the typical time scale of 
recollapsing for various cases in table 1. 
\begin{center}
\begin{tabular}{|c|c|c|c|} 
\hline
 & $\Omega_I>1$ & $\Omega_I=1$ & $\Omega_I<1$ \\ \hline
\hline
$\Lambda>0$ & inflation & inflation & inflation \\ \hline
$\Lambda=0$ (radiation dominated) & $O(t_I)(\Omega_I-1)^{-3/2}$ & $\infty$
 & $\infty$
\\ \hline
$\Lambda=0$ (matter dominated) & $O(t_I)(\Omega_I-1)^{-1}$ & $\infty$ & 
$\infty$ \\ \hline
$\Lambda<0$ & $(O(t_I))$ &$(O(t_I))$ & $O(t_I)$ \\ \hline
\end{tabular}
\vspace{0.2cm}
{\rm table 1. The lifetime for recollapsing}
\end{center}
Here $\Omega_I$ is the ratio of the density to the critical density
at the initial time $t_I$.
We can easily find several examples in which
some of bubbles are recollapsing and others are expanding. For example,
consider the following situation. The universe has the larger energy 
density than the
critical density at the time $t_{PT}$ 
when the phase transition occurs. Before the phase transition, the 
potential has only one minimum, but after the phase transition, the 
potential has one global minimum with vanishing cosmological 
constant and the other minimum which is local. The bubble universe
with the true vacuum recollapses because the energy density is 
larger than
the critical density. The time scale for the recollapsing is 
$O(t_I)$ unless the energy density is tuned to the critical density.
On the other hand, the bubble 
universe with the false 
vacuum inflates if the cosmological constant dominates the energy
density before reaching the time when the bubble begins to shrink.

In order to regard the inflating universe as our universe, the inflation
must stop some time because our universe does not inflate at present, 
namely cosmological constant must vanish till now. Here we expect that 
there is an unknown mechanism which realizes the vanishing cosmological 
constant. And such a mechanism automatically stop the inflation and the 
vacuum energy
is released in the universe(reheating). Of course we have no reliable 
mechanism for the vanishing cosmological constant, so we do not discuss
more on how to stop the inflation here.

It is interesting that the evolution of the universe prefers local minima
in many cases, in contrat with the common sense that the physics prefers
less energy state like the global minimum.
Actually the local minimum is unstable in the context of the ordinary 
field theory. The lifetime of the unstable vacuum is estimated by using 
instanton calculation which was discussed by Coleman
\cite{coleman}.
The condition for longer lifetime than the age of our universe 
($\sim 10^{20}$ seconds) gives
$10^{200}\frac{\Lambda^4}{(GeV)^4} \leq e^{S(\bar \phi)}$,
where $S(\phi)$ and $\Lambda$ are the action of the scalar $\phi$
and the typical scale of the potential, respectively. Here $\bar \phi$ 
is an instanton solution. 
When the difference of the potential energy between the two minima 
$\Delta V$ is much smaller than the typical scale $\Lambda^4$, 
then we get $S(\bar \phi)\sim 2 \pi^2 O(1)/\epsilon^3$,
where $\Delta V \sim \epsilon \Lambda^4$.
Therefore $\epsilon \sim 1/10$ is small enough to satisfy the above 
condition. This situation seems not to require strong fine tuning, or
rather it is not natural that a scalar 
potential has no local minimum with the above feature. Therefore we 
think that it is not unnatural that we are living in a false vacuum.

\section*{Primordial Black Holes}
If the recollapsing bubbles exist in our present horizon, these may 
become black holes. Here we make a simple argument for the bubbles with 
horizon size radius at the phase transition, because this is a typical 
scale in the context of cosmology.

The total mass of the bubble can be estimated from the size of the 
bubbles and the energy density. We have already taken the horizon 
length as the size of the bubbles. Since the bubble is recollapsing 
soon after the phase transition, the density is simply estimated 
by the phase transition temperature. Therefore in such case, the mass 
of the bubble becomes horizon mass at the phase transition;
$M_{Bubble}\sim d_H^3 \rho\sim M_P^3/T_{PT}^2$.
The Schwarzschild radius for the mass can be estimated as
$R_S\sim 1/H$.
It is surprising that the Schwarzschild radius is the same order of 
the magnitude of the horizon length.
Therefore we can expect that the bubbles will become black holes.
Notice that the density outside the bubbles becomes smaller than the 
inside density in contrast with the usual
situation in which the density of the universe is totally homogeneous
and there is no special region like black holes.

The black holes evaporate by the Hawking radiation. 
The lifetime is estimated as
$\tau_{BH}\sim 2560\pi G^2M_{BH}^3/g_*\sim 2560\pi
   M_P^5/(g_*T_{PT}^6)$.
Here $g_*$ is the number of freedom at the temperature.
As the summary, see table 2.
\begin{center}
\begin{tabular}{|c|c|c|c|c|c|} 
\hline
$T_{PT}$ ({\rm GeV}) & 1 & 100 & $10^9$ & $10^{12}$ & $10^{19}$ 
 \\ \hline
$M_{BH}$ (g)& $O(10^{33})\sim O(M_{Solar})$ & $O(10^{29})$ &
$O(10^{15})$  & $O(10^9)$ & $O(10^{-5})$  \\ \hline
$T_H({\rm GeV})$ & $O(10^{-21})$ & $O(10^{-17})$ & $O(10^{-3})$
 & $O(10^3)$ & $O(10^{17})$ \\ \hline
$\tau_{BH}$ (s) & $O(10^{74})$ & $O(10^{62})$ & $O(10^{20})\sim \tau_0$
 & $O(1)$ & $O(10^{-42})$ 
 \\ \hline
\end{tabular}

\vspace{2mm}
table 2. Here $M_{Solar}$ is the solar mass and $\tau_0$ is the age of
 our universe.
\end{center}
When the temperature of the phase transition is smaller than $10^9$ GeV,
we may find the small black holes in our universe.
Actually Massive Astrophysical Compact Halo Object(MACHO)s with the 
solar mass were 
discovered by gravity lensing effect
\cite{MACHO}. It is interesting that
the mass scale is related with the QCD phase transition.
When the temperature of the phase transition is larger than $10^9$ GeV,
the black holes evaporate until now. Global minima disappear in the 
history and only the universe with the local minima is realized.

\section*{Cosmological SUSY Breaking}
If SUSY exists, the scalar potential also exists. Generally the scalar 
potential has local minima with longer life than the age of our 
universe. Since the cosmology prefers the local minima, it seems to be 
natural that the SUSY is spontaneously broken cosmologically even if 
the scalar potential has SUSY vacua. We call this phenomenon 
cosmological SUSY breaking.

The natural scale will be the scale of the typical scale of the 
potential. If SUSY exists at the Planck scale, the Planck scale is 
the natural scale for the SUSY breaking.
Of course our argument is strongly dependent on the initial conditions, 
so we can consider the case of lower SUSY breaking scale. By using 
cosmological SUSY breaking, we can make a simpler model
\cite{maekawa}
 in which SUSY is cosmologically broken and the SUSY breaking is 
 mediated by the standard gauge interactions. The point
is that these models are allowed to have SUSY vacua.

\end{document}